\documentclass[aps,twocolumn,prb,reprint,floatfix,showpacs,lengthcheck,citeautoscript,superscriptaddress,nofootinbib]{revtex4-2}
\usepackage{amsmath}
\usepackage{amssymb}
\usepackage{graphicx}
\usepackage{bm}
\usepackage{float}
\usepackage{xcolor}
\usepackage{times}
\usepackage{notes2bib}
\usepackage{verbatim}
\usepackage[none]{hyphenat}
\usepackage[colorlinks=true,
linkcolor=blue, 
citecolor=blue, 
urlcolor=blue,
bookmarks=true,
hypertexnames=true]{hyperref}

\newcommand{\e}{\epsilon}
\newcommand{\ci}{\mathrm{i}}

\newcommand{\ket}[1]{| #1 \rangle}
\newcommand{\bra}[1]{\langle #1 |}
\newcommand{\bigdot}{\boldsymbol{\cdot}}

\newcommand{\hw}{\hbar\omega}
\newcommand{\w}{\omega}
\newcommand{\moi}{\leqslant}
\newcommand{\dk}{\int\,\hspace{-2mm}\frac{d\bm{k}}{(2\pi)^3}\,}

\newcommand{\bk}{\bm{k}}
\newcommand{\br}{\bm{r}}

\newcommand{\bG}{\bm{G}}
\newcommand{\bp}{\bm{p}}

\newcommand{\p}{\partial}

\newcommand{\chil}{\chi_{ij}}

\newcommand{\ddr}{\frac{\partial}{\partial \bm{r}}}
\newcommand{\ddk}{\frac{\partial}{\partial \bm{k}}}

\begin{document}
	\title{The Thomas-Reiche-Kuhn sum rule as a consequence of a non-singular \\optical susceptibility in semiconductors}
	\author{Angiolo Huamán}
	\affiliation{Independent Researcher, Fayetteville, Arkansas 72701, USA}
	\begin{abstract}
		The Thomas-Reiche-Kuhn (TRK)   sum rule constrains the optical susceptibility  by relating it to the electron density of the material, and it has conventionally been derived by combining the Kramers-Kronig relations with the high frequency limit of the   linear optical susceptibility tensor $\chil$. While these are arguments based on causality and quasiclassical behavior, respectively, no derivation using  the quantum mechanical state of the material is widely known.  Here we close this gap by showing that a proof of the TRK sum rule relying only on the Bloch eigenstates of the solid is possible. To do this, we use the well-known procedure to lift the singularity at zero frequency in the optical susceptibility tensor in semiconductors, and show that this regularization  naturally leads to the TRK sum rule.  To reach this result, we first present a thorough description of the momentum matrix elements between Bloch eigenfunctions bypassing the so-called $\bm{k}-$representation. This proof highlights the intrinsically microscopic and quantum mechanical origin of the optical properties of a solid.
	\end{abstract}
	\date{\today}	\maketitle
\section{Introduction}
The optical properties of a material, that is, its response to an applied time-dependent field, are of fundamental importance because their connection to the energy bands and wavefunctions serves as a probe of the electronic properties of quantum systems. In fact, before the advent of more modern techniques like ARPES~\cite{Sobota2021}, much information about the electronic states of a system were inferred from its optical response to an applied optical external field. 
 
On the theoretical realm, the Thomas-Reiche-Kuhn (TRK) sum rule~\cite{Barnett2012} for the optical susceptibility of solids remains a crucial tool for validating calculations of optical susceptibilities. Similar to the sum rules in atoms~\cite{BransdenPAM}, the optical response of a periodic solid, that is, its frequency-dependent optical susceptibility $\chil(\w)$, is constrained by the TRK sum rule, which relates the imaginary part of $\chil$ to the electron density $n_e$ of the material. In Gaussian units, this relation reads:
\begin{equation}\label{eq:TRK}
	\int_0^\infty d\w\, \w \,\text{Im}[\chil]=\frac{ \pi e^2}{2m}n_e,
\end{equation}
where $\w$ is the light frequency, $m$ and $-e$ are the electron mass and charge, respectively. Equation~\eqref{eq:TRK} is customarily proved by a combination of the Kramers-Kronig relations (which connect the real and imaginary parts of $\chil$) and the high-frequency limit of  $\chil$, that is obtained on purely physical grounds~\cite{LandauEOCM,CallawayQTOTSS}, and that is valid both for insulators  and metals, showcasing the fact that at sufficiently high frequencies, there are not even quantitative differences between these two kinds of materials. The same procedure leads to similar sum rules for the nonlinear optical susceptibility~\cite{Bassani1991,Koutserimpas2025}. 

On another note, in the independent particle approximation (IPA), we can obtain expressions for the optical susceptibility, both linear and nonlinear, by solving the equation of motion of the density matrix in a perturbative way~\cite{Cabellos2009}. The resulting equations are written in terms of the Bloch eigenstates and energy bands of the underlying material, but have the disadvantage of being singular at zero frequency although it can be shown that this is a mathematical rather than a physical feature: by using the properties of the Bloch eigenstates, we can show that the terms in $\chil$ giving a singular behavior are in fact zero.

The purpose of this article is a fundamentally theoretical one: in the case of semiconductors (where the singularity at $\w=0$ is merely mathematical), Eq.~\eqref{eq:TRK} can be obtained purely from the properties of the Bloch eigenfunctions and their momentum matrix elements (MMEs), and from the condition that the susceptibility has to be non-singular at $\w=0$. This approach reinforces a fundamental argument in condensed matter theory: that all physical properties of a material must be rooted in its quantum mechanical state. A key step in doing this is to reduce the singular factor to an integral of the effective mass tensor over the Brillouin zone (BZ).  In doing this, we prefer to avoid the so-called $\bk-$representation, and calculate the MMEs directly from the symmetry properties of the Bloch eigenstates. This regularization of $\chil$ is well known but, to the best of our knowledge, its connection to the TRK sum rule has not been sufficiently investigated.

The rest of the article is organized as follows. In Sec.~\ref{sec:normalization} we present the way Bloch wavefunctions are normalized. Section~\ref{sec:MMEs} deals with the calculation of the MMEs and the effective mass tensor, in a form suitable for the study of optical properties. Our main result, the TRK sum rule from the properties of the Bloch waves is presented in Sec.~\ref{sec:TRK}. The conclusions we arrived at are summarized in Sec.~\ref{sec:conclusions}.

\section{Normalization of a Bloch function.}\label{sec:normalization}
Let us start by recalling some basic properties of a Bloch wave. In a perfect crystal of volume $V$ and lattice vectors $\bm{a}_j$ ($j=1,2,3$), the  electrons move in a periodic potential $V(\br)$ satisfying $V(\br)=V(\br+\bm{R})$, where $\bm{R}$ is a point in the lattice, $\bm{R}=\sum_j n_j\bm{a}_j$, with $n_j$ integers. The eigenstates of the crystal Hamiltonian $H_0=\bp^2/2m+V(\br)$ are Bloch waves $\psi_{n\bk}(\br)=e^{\ci\bk\bigdot\br}u_{n\bk}(\br)$ satisfying $H_0\psi_{n\bk}=\e_{n\bk}\psi_{n\bk}$, where $u_{n\bm{k}}(\bm{r})$ has the periodicity of the crystal potential: $u_{n\bm{k}}(\bm{r})=u_{n\bm{k}}(\bm{r}+\bm{R})$.  If the crystal has $N_j$ unit cells along the $\bm{a}_j$ lattice vector, periodic or Born-von Karman boundary conditions~\cite{Ashcroft76} state that the $\bm{k}$ vector can only have the values:
\begin{equation}\label{eq:vK}
	\bm{k}=\sum_{j=1}^3 \frac{n_j}{N_j}\bm{b}_j,
\end{equation}
where $n_j$ are integers ($n_j=1,\cdots,N_j$) and $\bm{b}_j$ are the reciprocal vectors. In the limit of a very large sample, where $\bk$ can be taken as a continuous vector variable, the Bloch eigenstates are normalized according to:
\begin{equation}\label{eq:norm}
	\int_V d\bm{r}\, \psi_{n'\bk'}^*\psi_{n\bk}=\delta(\bk'-\bk)\delta_{n'n},
\end{equation}
which can be established by dividing the volume $V$ into $N$ ($N\sim N_1N_2N_3$) primitive cells of volume $\Omega$ and noting that $\sum_{\bm{R}}e^{i(\bm{k}-\bm{k}')\bigdot\bm{R}}=\frac{(2\pi)^3}{\Omega}\delta(\bm{k}-\bm{k}')$ in the limit of a very large solid [under the assumption in Eq.~\eqref{eq:vK}]. This imposes the following normalization on $u_{n\bk}$:
\begin{equation}\label{eq:u_norm}
\int_{\text{uc}} d\bm{r}\,u_{n'\bk}^*u_{n\bk}=\frac{\Omega}{(2\pi)^3}\delta_{n'n},
\end{equation}
(the integral is taken over the unit cell, {\it uc}) 
according to which $u_{n\bk}$ is dimensionless.  It is useful to point out that, since the normalization in Eq.~\eqref{eq:norm} is not dimensionless, the matrix elements between Bloch eigenstates will have an additional dimension $\sim |\bk|^3$. In the following sections, we will see that dimensionally correct matrix elements can be defined at the {\it same} $\bk$ point.

\section{The matrix elements of the momentum operator}\label{sec:MMEs}
Because the Bloch eigenfunctions are not bounded, the calculation of matrix elements between them presents particular difficulties. In the case of the momentum operator, it entails the appearance of Dirac deltas in the quasimomentum $\bk$. Before going into a detailed presentation, let us state the most general form of these matrix elements. By definition, they are equal to:
\begin{equation}
	\bra{n'\bm{k}'} \bm{p} \ket{n\bm{k}} = \int_V d\bm{r}\,\psi_{n'\bk'}^*\frac{\hbar}{\ci}\ddr\psi_{n\bk},
\end{equation}
where the integral is taken over the volume $V$ of the solid and $\ket{n\bk}$ is simply the Dirac notation for the eigenstate $\psi_{n\bk}$. By replacing $\psi_{n\bk}=e^{\ci\bk\bigdot\br}u_{n\bk}$ in the equation above and after reducing it to the unit cell, we get:
\begin{align}\label{eq:p_av}
	\bra{n'\bm{k}'} \bm{p} \ket{n\bk}=&\delta(\bk-\bk')\Big[ \hbar\bk\,\delta_{nn'}+\notag\\
	&\frac{(2\pi)^3}{\Omega}\int_{\text{uc}} d\bm{r}\,u^*_{n'\bk}\frac{\hbar}{\ci}\ddr u_{n\bk} \Big].
\end{align}
This result is a statement of the important fact that the momentum operator is diagonal in the quasimomentum $\bk$, and it allows for a natural definition of the MMEs at the {\it same} $\bk$ point, $\bm{p}_{n'n}(\bk)$, such that $\bra{n'\bm{k}'} \bm{p} \ket{n\bk}=\delta(\bk-\bk')\bm{p}_{n'n}(\bk)$:
\begin{equation}\label{eq:p_samek}
	\bm{p}_{n'n}= \hbar\bk\,\delta_{n'n}+
	\frac{(2\pi)^3}{\Omega}\int_{\text{uc}} d\bm{r}\,u^*_{n'\bk}\frac{\hbar}{\ci}\ddr u_{n\bk},
\end{equation}
which does have units of momentum. Also, because of time-reversal symmetry we can choose $u_{n\bk}$ in such a way that it satisfies $u_{n\bk}^*=u_{n,-\bk}$, thus from Eq.~\eqref{eq:p_samek} we have $\bp_{n'n}(-\bk)=-\bp_{n'n}^*(\bk)$.

Before proceeding further, a comment is necessary. In the coupling with an external oscillating electric field, energy and quasimomentum must be conserved. In the latter case, we should have $\bk_f=\bk_i+\bm{K}$, where $\bk_i$ ($\bk_f$) is the initial (final) electron quasimomentum, while $\bm{K}$ is that of the photon. However, in the optical range $\hw\simeq 1\,$eV, $|\bm{K}|$ is much smaller than $|\bk_i|$ (of the order of $10^{-4}$\AA$^{-1}$), so that we can safely assume that $\bk_i\simeq\bk_f$ upon absorbing a photon. Mathematically, this amount to including only MMEs with the same $\bk$, as those defined in Eq.~\eqref{eq:p_samek}. Having this is mind, from now on we will focus on this kind of MMEs only.

\subsection{Momentum matrix elements based on the Fourier expansion of the Bloch eigenfunctions. }

Most calculations involving optical properties of solids are written in terms of the matrix elements of the velocity operator $\bm{v}$ rather than the momentum ones, the former of which is given by:
\begin{equation}\label{eq:v}
	\bm{v}=\frac{1}{\ci\hbar}[\bm{r},H].
\end{equation}
In the absence of illumination, $H=H_0=\bm{p}^2/2m + V(\bm{r})$, and the usual result $\bm{v}=\bm{p}/m$ is recovered, where $m$ is the electron mass. Using Eq.~\eqref{eq:v} we can express the momentum matrix elements in terms of energy differences and other quantities related to the periodicity of $u_{n\bk}$, but this requires the introduction of the matrix elements of the position operator $\br$. The determination of these matrix elements is not straightforward, although has been discussed abundantly. This calculation relies on the so-called $\bk-$representation, where operators act upon the $\bk-$dependent coefficients  of a wavefunction constructed as a combination of Bloch eigenstates~\cite{BLOUNT1962305,LandauSP2}.

Although this approach should be considered standard by now, the same expressions for the MMEs can be obtained by insisting with the usual $\br-$representation, using the periodicity of the functions $u_{n\bk}(\br)$~\cite{KittelQTS}. We take this path to a full extent and use it to obtain expressions for the MMEs and the effective mass tensor, as they are often used in articles on first principles optics in semiconductors.

The functions $u_{n\bk}(\br)$, having the periodicity of the lattice, can be written as a Fourier expansion:
\begin{equation}\label{eq:u_G}
	u_{n\bk}(\br)=\sum_{\bG} e^{\ci\bG\bigdot\br}f_{n\bG}(\bk),
\end{equation}
where the sum goes over the lattice vector $\bG$ of reciprocal space, $\bG=\sum_j n_j \bm{b}_j$, $j=1,2,3$ and $n_j$ integers. The normalization in Eq.~\eqref{eq:u_norm} can be written using Eq.~\eqref{eq:u_G} and results in:
\begin{equation}\label{eq:f_norm}
	\sum_{\bG}f_{n'\bG}^*f_{n\bG}=\delta_{n'n}/(2\pi)^3,
\end{equation}
where we have used $\int_{uc}d\br\,e^{\ci\br\bigdot(\bG-\bG')}=\Omega\,\delta_{\bG,\bG'}$. Replacing Eq.~\eqref{eq:u_G} into $\psi_{n\bk}=e^{\ci\bk\bigdot\br}u_{n\bk}$ and using the Schrödinger equation $[\bp^2/2m+V(\br)]\psi_{n\bk}=\e_{n\bk}\psi_{n\bk}$, we obtain an algebraic equation for the $f_{n\bG}$ coefficients:
\begin{equation}\label{eq:f_eq}
	\frac{\hbar^2}{2m}|\bk+\bG|^2f_{n\bG}+\sum_{\bm{g}}V(\bG-\bm{g})f_{n\bm{g}}=\e_{n\bk}f_{n\bG}.
\end{equation}
In this equation, $V(\bm{q})$ is the Fourier component $\frac{1}{\Omega}\int_{uc}d\br\, e^{-\ci\bm{q}\bigdot\br}V(\br)=V(\bm{q})$ of the crystal potential $V(\br)$. Also, replacing Eq.~\eqref{eq:u_G} into Eq.~\eqref{eq:p_samek}, we get the following expression for the MMEs at the same $\bk$:
\begin{equation}\label{eq:p_gen}
	\bp_{n'n}=(2\pi)^3\sum_{\bG}\hbar(\bk+\bG)f_{n'\bG}^*f_{n\bG}.
\end{equation}
It is customary, and also useful, to write the equation above by separating the cases $n'=n$ and $n'\neq n$. For one thing, they have pretty different forms when written in terms of the energy bands, as we will show next.  In the former case, we have:
\begin{equation}\label{eq:p_nn}
	\bp_{nn}=(2\pi)^3\sum_{\bG}\hbar(\bk+\bG)|f_{n\bG}|^2,
\end{equation}
an expression that will be shown to depend only on the $\bk-$derivatives of the energy bands $\e_{n\bk}$. To show this, we take the $\bk-$derivative of Eq.~\eqref{eq:f_eq}, to get:
\begin{align}\label{eq:f_eq1}
	&\frac{\hbar^2}{m}(\bk+\bG)f_{n\bG}+\frac{\hbar^2}{2m}|\bk+\bG|^2\ddk f_{n\bG}+\notag\\
	&\sum_{\bm{g}}V(\bG-\bm{g})\ddk f_{n\bm{g}}=f_{n\bG}\,\ddk \e_{n\bk}+\e_{n\bk}\ddk f_{n\bG}.
\end{align}
By multiplying the equation above by $f_{n\bG}^*$ and summing over $\bG$, we get after some rearrangement:
\begin{widetext}
\begin{equation}\label{eq:f_dev}
	\frac{\hbar^2}{m}\sum_{\bG} (\bk+\bG)|f_{n\bG}|^2-\ddk\e_{n\bk}\sum_{\bG}|f_{n\bG}|^2=\sum_{\bG}\big[\e_{n\bk}-\frac{\hbar^2}{2m}|\bk+\bG|^2\big]f_{n\bG}^*\ddk f_{n\bG}-\sum_{\bG,\bm{g}}V(\bG-\bm{g})f_{n\bG}^*\ddk f_{n\bm{g}}.
\end{equation}
\end{widetext}
Let us now show that the right-hand side (rhs) in the equation above is zero. To do this, we take the complex conjugate of Eq.~\eqref{eq:f_eq}, multiply it by $\ddk f_{n\bG}$ and sum it up over $\bG$:
\begin{align}
	\frac{\hbar^2}{2m}|\bk+\bG|^2f_{n\bG}^*\ddk f_{n\bG}+&\sum_{\bG,\bm{g}}f_{n\bm{g}}^*V^*(\bG-\bm{g})\ddk f_{n\bG}=\notag\\
	&\e_{n\bk}\sum_{\bG}f_{n\bG}^*\ddk f_{n\bG}.
\end{align}
After swapping dummy indices $\bG$ and $\bm{g}$ in the second term and noticing that $V^*(\bm{g}-\bG)=V(\bG-\bm{g})$ [because $V(\br)$ is a real field], we get the expected result. Thus, Eq.~\eqref{eq:f_dev} goes over into:
\begin{equation}\label{eq:ek_der}
	\frac{\hbar^2}{m}\sum_{\bG} (\bk+\bG)|f_{n\bG}|^2=\ddk\e_{n\bk}\times \sum_{\bG}|f_{n\bG}|^2.
\end{equation}
By using the normalization condition Eq.~\eqref{eq:f_norm} and replacing into Eq.~\eqref{eq:p_nn}, we finally get:
\begin{equation}\label{eq:pnn}
\bp_{nn}=\frac{m}{\hbar}\ddk \e_{n\bk},	
\end{equation}
 from which it follows the well known expression $(1/\hbar) \ddk \e_{n\bk}$ for the electron velocity in a band. 

For $n'\neq n$ (the so-called {\it interband} matrix elements or {\it transitions}), we have from Eq.~\eqref{eq:p_gen}:
\begin{equation}\label{eq:p_nm}
	\bp_{n'n}=(2\pi)^3\sum_{\bG}\hbar\bG f_{n'\bG}^*f_{n\bG},
\end{equation}
the term with $\bk$ vanishes because it is accompanied by $\sum_{\bG} f_{n'\bG}^*f_{n\bG}$, which is zero  [$n'\neq n$, see Eq.~\eqref{eq:f_norm}]. Our goal is to introduce in the equation above  the energy bands. To do this, let us notice that since $u_{n\bk}$ is spatially periodic, its $\bk-$derivatives are also so, and thus $\ddk u_{n\bk}$ can be written as a combination of $u_{n'\bk}$:
\begin{equation}
	\ddk u_{n\bk}(\br)=-\ci\sum_{n'} \bm{\mathcal{M}}_{n'n}(\bk) u_{n'\bk}(\br)
\end{equation}
The factor $-\ci$ is included to make $\bm{\mathcal{M}}_{n'n}$ Hermitian (see below). The $\bk-$dependent factors $\bm{\mathcal{M}}_{n'n}(\bk)$ are related to the coefficients used in defining the matrix elements of the position operator~\cite{BLOUNT1962305,LandauSP2}. After using Eq.~\eqref{eq:u_G} in both sides of the equation above and equating the coefficients of $e^{\ci\bG\bigdot\br}$, we get the following relation between $\bm{\mathcal{M}}_{n'n}(\bk)$ and the coefficients $f_{n\bG}$:
\begin{equation}\label{eq:Ms}
	\ddk f_{n\bG}=-\ci\sum_{n'} \bm{\mathcal{M}}_{n'n} f_{n'\bG}.
\end{equation}
From this equation and Eq.~\eqref{eq:f_norm} we have:
\begin{equation}
	\bm{\mathcal{M}}_{n'n}=\ci(2\pi)^3\sum_{\bG} f_{n'\bG}^*\ddk f_{n\bG}.
\end{equation}
Moreover, by taking the $\bk-$derivative of Eq.~\eqref{eq:f_norm} it follows that $\bm{\mathcal{M}}_{n'n}$ is Hermitian: $\bm{\mathcal{M}}_{n'n}^*=\bm{\mathcal{M}}_{nn'}$. 

Let us obtain an expression for $\bp_{n'n}$ ($n'\neq n$) as a function of $\bm{\mathcal{M}}_{n'n}$ and the energy bands $\e_{n\bk}$. We start again with Eq.~\eqref{eq:f_eq1}. After multiplying it by $f_{n'\bG}^*$ ($n'\neq n$) and summing over $\bG$ we get:
\begin{widetext}
\begin{equation}\label{eq:f_dev2}
	\frac{\hbar^2}{m}\sum_{\bG} \bG f_{n'\bG}^*f_{n\bG}-\e_{n\bk}\sum_{\bG}f_{n'\bG}^*\ddk f_{n\bG}=-\frac{\hbar^2}{2m}\sum_{\bG}|\bk+\bG|^2 f_{n'\bG}^*\ddk f_{n\bG}-\sum_{\bG,\bm{g}}V(\bG-\bm{g})f_{n'\bG}^*\ddk f_{n\bm{g}},
\end{equation}
\end{widetext}
where we have used again $\sum_{\bG}f_{n'\bG}^*f_{n\bG}=0$ because $n'\neq n$. Let us work out the rhs. We notice that if we start from Eq.~\eqref{eq:f_eq} but for $f_{n'\bG}$ instead, take the complex conjugate, multiply it by $\ddk f_{n\bG}$ and summing over $\bG$ we obtain:
\begin{align}
	\frac{\hbar^2}{2m}\sum_{\bG} |\bk+\bG|^2f_{n'\bG}^*\ddk f_{n\bG}+&\sum_{\bG,\bm{g}}V^*(\bG-\bm{g})f_{n'\bm{g}}^*\ddk f_{n\bG}=\notag\\&\e_{n'\bk}\sum_{\bG}f^*_{n'\bG}\ddk f_{n\bG}.
\end{align}
After swapping $\bm{g}$ and $\bG$ dummy indices the equation above [and using again $V^*(\bm{g}-\bG)=V(\bG-\bm{g})$] we can replace the result in Eq.~\eqref{eq:f_dev2} to obtain:
\begin{equation}
	\frac{\hbar^2}{m}\sum_{\bG}\bG\,f_{n'\bG}^*f_{n\bG}=\e_{nn'\bk}\sum_{\bG}f_{n'\bG}^*\ddk f_{n\bG},
\end{equation}
where the notation $\e_{nn'\bk}=\e_{n\bk}-\e_{n'\bk}$ has been introduced. By replacing $\ddk f_{n\bG}$ with the expression in Eq.~\eqref{eq:Ms} and using again Eq.~\eqref{eq:f_norm} we finally have:
\begin{equation}\label{eq:p_nm_final}
	\bp_{n'n}=\ci\frac{m}{\hbar}\e_{n'n\bk}\bm{\mathcal{M}}_{n'n},\hspace{3mm}(n'\neq n).
\end{equation}
Equations~\eqref{eq:pnn} and~\eqref{eq:p_nm_final} are the ones most commonly obtained using the $\bk-$representation and  will prove themselves fundamental to obtain the effective mass tensor, as we do in the following section.

\subsection{The effective mass tensor and its relations with the momentum matrix elements.}
An essential ingredient to obtain the Thomas-Reiche-Kuhn sum rule is the effective mass tensor. It is well known that it can be written as a second derivative of the energy bands $\e_{n\bk}$~\cite{Ashcroft76,LandauSP2}, but it is its relation with the momentum matrix elements what matters most for our goals, so we set up to obtain this expression. By separating the $i$-th Cartesian components of Eq.~\eqref{eq:ek_der} we have~\cite{KittelQTS}:
\begin{equation}
	\p_i\e_{n\bk}=\frac{\hbar^2}{m}k_i+(2\pi)^3\frac{\hbar^2}{m}\sum_{\bG}G_i|f_{n\bG}|^2,
\end{equation}
where $\p_i=\p/\p k_i$, and the Cartesian component of $\bk$ and $\bG$ are $k_i$ and $G_i$, respectively ($i=1,2,3$). We have also used $\sum_{\bG} |f_{n\bG}|^2=1/(2\pi)^3$. Taking an extra derivative $\p_j=\p/\p k_j$, we have:
\begin{equation}\label{eq:m_inv1}
	\p^2_{ij}\e_{n\bk}=\frac{\hbar^2}{m}\delta_{ij}+2(2\pi)^3\frac{\hbar^2}{m}\text{Re}\Big[\sum_{\bG}G_i\p_jf_{n\bG}^* f_{n\bG}\Big],
\end{equation}
where Re[$\cdots$] stands for the real part. Using the $j-$th component of Eq.~\eqref{eq:Ms} we have:
\begin{align}\label{eq:aux}
&\sum_{\bG}G_i\p_jf_{n\bG}^* f_{n\bG}= \ci \sum_{n'}\mathcal{M}^*_{j;n'n}\sum_{\bG} G_i f^*_{n'\bG}f_{n\bG}=\notag\\
	&\ci \mathcal{M}^*_{j;nn}\sum_{\bG} G_i |f_{n\bG}|^2+\notag\\
	&\hspace{10mm}\ci \sum_{\substack{n'\\n'\neq n}}\mathcal{M}^*_{j;n'n}\sum_{\bG} G_i f^*_{n'\bG}f_{n\bG},
\end{align}
where we have isolated the term $n'=n$ in the sum over $n'$. It is clear that this term is a pure imaginary number, as the sum over $\bG$ is real and $\mathcal{M}^*_{j;nn}=\mathcal{M}_{j;nn}$, as the matrix $\bm{\mathcal{M}}_{n'n}$ is Hermitian ($\mathcal{M}_{j;n'n}$ is the $j-$th component of $\bm{\mathcal{M}}_{n'n}$), then it will not contribute to the real part in Eq.~\eqref{eq:m_inv1}, and we can ignore it from now on. We then define $\Pi_{ij;n}$:
\begin{equation}\label{eq:Pi}
	\Pi_{ij;n}=\ci \sum_{\substack{n'\\n'\neq n}}\mathcal{M}^*_{j;n'n}\sum_{\bG} G_i f^*_{n'\bG}f_{n\bG}.
\end{equation}
With this, Eq.~\eqref{eq:m_inv1} can be written as follows:
\begin{equation}\label{eq:m_inv2}
	\p^2_{ij}\e_{n\bk}=\frac{\hbar^2}{m}\delta_{ij}+(2\pi)^3\frac{\hbar^2}{m}[\Pi_{ij;n}+\Pi_{ij;n}^*].
\end{equation}
 From Eq.~\eqref{eq:p_nm}, the sum $\sum_{\bG} G_i f^*_{n'\bG}f_{n\bG}$ in $\Pi_{ij;n}$ is equal to $p_{i;n'n}/(\hbar(2\pi)^3)$,  $p_{i;n'n}$ is the $i-$th component of $\bp_{n'n}$. Thus, we have in Eq.~\eqref{eq:Pi}:
\begin{equation}
	\Pi_{ij;n}=\frac{\ci}{\hbar(2\pi)^3}\sum_{\substack{n'\\n'\neq n}}\mathcal{M}^*_{j;n'n} p_{i;n'n}.
\end{equation}
By replacing $\mathcal{M}^*_{j;n'n}$ using Eq.~\eqref{eq:p_nm_final} in the expression above  we obtain:
\begin{equation}
	\Pi_{ij;n}=\frac{1}{m(2\pi)^3}\sum_{\substack{n'\\n'\neq n}}\frac{p_{j;nn'}p_{i;n'n}}{\e_{nn'\bk}}.
\end{equation}
In getting this we have used $p^*_{j;n'n}=p_{j;nn'}$, as the momentum operator is Hermitian. Replacing this expression and its complex conjugate in Eq.~\eqref{eq:m_inv2} we have, after dividing by $1/\hbar^2$, (and omitting the $\bk$ dependence for brevity):
\begin{align}\label{eq:eff_m}
	\frac{1}{\hbar^2}\p^2_{ij}\e_{n\bk}=\frac{\delta_{ij}}{m}+\frac{1}{m^2}\hspace{-1mm}&\sum_{\substack{n'\\ n'\neq n}}\hspace{-1mm}\frac{1}{\e_{nn'}}[p_{j;nn'}p_{i;n'n}+\notag\\
	&p_{i;nn'}p_{j;n'n}]=\Big(\frac{1}{m^*}\Big)_{ij},
\end{align}
which is the expression for the effective mass tensor $(1/m^*)_{ij}$ most commonly used in papers on optics. The fact that it is written as a  second derivative of the energy bands will be crucial for proving the Thomas-Reiche-Kuhn sum rule, as we will do in the next sections. This finalizes our study of the MMEs between Bloch eigenstates. In the rest of this article we will pass to the actual problem of deriving the TRK sum rule in semiconductors.

\section{The TRK sum rule for the linear optical susceptibility.}\label{sec:TRK}
In semiconductor materials, it can be shown~\cite{Cabellos2009} that after a direct calculation of the linear optical susceptibility tensor $\chil$---by using time-dependent perturbation theory on the density matrix---it exhibits a divergence at zero frequency, which is in contradiction with the fact that insulating materials have a finite static ({\it i.e.}, $\w\rightarrow 0$) susceptibility. However, in semiconductors this divergence is artificial: it can be shown (see next subsection) that by a suitable manipulation this singularity can be removed, leaving an expression that is regular at $\w=0$. It is less widely known, however, that this same procedure  naturally leads to the Thomas-Reiche-Kuhn sum rule, as we show in what follows.

\subsection{Non-singular linear optical susceptibility.}
Using time-dependent perturbation theory in the matrix density we can get an expression for the linear optical susceptibility by first obtaining the induced current, from which we can determine the induced polarization~\cite{JacksonCE}. There exist many levels of approximation in getting the optical susceptibility: we will assume that the field acting on the material can be treated classically (other formalisms explicitly quantize the electromagnetic field~\cite{Garcia2023}); and that it is monochromatic and spatially homogeneous (spatial variations are usually of no effect, although they can be sometimes important~\cite{LandauEOCM}). The assumption of a classical field restricts the validity of these results to not too weak field intensities, with a density of photons large enough so that it can be treated as a continuous variable. Also, we will work within the IPA: the mutual Coulomb interaction between electrons will be taken as {\it averaged out} and included in the crystal potential $V(\br)$.  With these assumptions, we have the following expression for the optical susceptibility tensor in Gaussian units~\cite{Cabellos2009}:
\begin{align}\label{eq:chil}
	\chil^S=-\frac{e^2}{m^2\w^2}\dk &\Big(\sum_{n,n'}m f_n\delta_{nn'}\delta_{ij}+\notag\\
	&\sum_{\substack{n,n' \\ n\neq n'}}(f_n-f_{n'})\frac{p_{j;nn'}p_{i;n'n}}{\e_{nn'}-\hw} \Big),
\end{align}
where $\w$ is the frequency of the external field (assumed having a small  positive imaginary part, $\w\equiv \w+i\eta$, $\eta>0$), $-e$ is the electron charge, and $f_n$ is the occupancy of the $\psi_{n\bk}$ Bloch eigenstate: $f_n=1$ ($f_n=0$) for valence (conduction) states. These occupations correspond to a semiconductor in equilibrium at zero temperature. It is also assumed that the optical field does not drive the material out of equilibrium in any important way, so that a perturbative approach is still valid. For simplicity, the dependence on $\bk$ has been omitted in the writing. The superscript $S$ indicates that the susceptibility, as it stands in Eq.~\eqref{eq:chil}, is {\it singular} at $\w=0$. This singularity is unphysical in semiconductors and insulators, whereas the situation is entirely different in metals~\cite{ZimanPTS}, in which the limit $\w\rightarrow0$ should correspond to the quasi-static regime where the electric displacement $\bm{D}(t)$ and the electric field $\bm{E}(t)$ are related by $\partial\bm{D}/\partial t=4\pi\sigma \bm{E}$ ($\sigma$ is the nearly constant electric conductivity), resulting in a permittivity proportional to $1/\w$~\cite{LandauEOCM}. This is the main reason this approach does not apply to that kind of materials. 

Although the removal of this singularity  has been shown elsewhere~\cite{Sipe1993},  we will repeat it here because of its connection to the TRK sum rule.  If we write $\e_{nn'}=\hbar\w_{nn'}$, we can use the partial fraction decomposition:
\begin{equation}
	\frac{1}{\w^2(\w_{nm}-\w)}=\frac{1}{\w\w_{nm}^2}+\frac{1}{\w^2\w_{nm}}+\frac{1}{\w_{nm}^2(\w_{nm}-\w)},
\end{equation}
which allows to write Eq.~\eqref{eq:chil} in the following fashion:
\begin{equation}\label{eq:chi_S}
	\chil^S=\frac{\chil^{(a)}}{\w}+\frac{\chil^{(b)}}{\w^2}+\chil,
\end{equation}
with:
\begin{align}
	\chil^{(a)}&=-\frac{e^2\hbar}{m^2}\hspace{-1mm}\dk\hspace{-3mm} \sum_{\substack{n,n'\\(n\neq n')}} \hspace{-2mm} \frac{f_{nn'}}{\e_{nn'}^2}p_{j;nn'}p_{i;n'n}, \label{eq:chi1} \\
	\chil^{(b)}&=-\frac{e^2}{m^2}\dk\Big(  \sum_{n,n'}m f_n\delta_{nn'}\delta_{ij}+\notag\\
	&\hspace{10mm}\sum_{\substack{n,n'\\(n\neq n')}}\frac{f_n-f_{n'}}{\e_{nn'}}p_{j;nn'}p_{i;n'n}\Big), \label{eq:chi2}\\
	\chil &=-\frac{e^2\hbar^2}{m^2}\hspace{-1mm}\dk\hspace{-2mm} \sum_{\substack{n,n'\\(n\neq n')}} \hspace{-2mm}  \frac{f_{nn'} p_{j;nn'}p_{i;n'n}}{\e^2_{nn'}(\e_{nn'}-\hw)},\label{eq:chi3}
\end{align}
the restriction $n\neq n'$ in Eqs.~\eqref{eq:chi1} and~\eqref{eq:chi3} has been added by virtue of the factor $f_{nn'}$ in both equations. The term  $\chil^{(a)}$, accompanying the factor $1/\w$ in $\chil^S$, vanishes because of time-reversal symmetry: $\bp_{nn'}(\bk)=-\bp^*_{nn'}(-\bk)$. This can be seen by separating the $\bk$-integral into two equal parts and doing the substitution $\bk\rightarrow-\bk$ in the second one. The term $\chil^{(b)}$, that would give the divergence $1/\w^2$,  can be shown to be zero by the following procedure~\cite{Ghahramani1991}. By separating the term $f_n-f_{n'}$ in the second term and swapping the dummy indices $n$ and $n'$ in the resulting equation, we have the following:
\begin{align}\label{eq:aux3}
	\chi_{ij}^{(b)}&=-e^2\dk \sum_n f_n\Big(\frac{\delta_{ij}}{m} + \notag\\
	&\frac{1}{m^2}\hspace{-2mm}\sum_{\substack{n'\\(n'\neq n)}}\hspace{-2mm}\frac{1}{\e_{nn'}}(p_{j;nn'}p_{i;n'n}+p_{i;nn'}p_{j;n'n})\Big),
\end{align}
which has the form of the effective mass tensor, Eq.~\eqref{eq:eff_m}:
\begin{equation}\label{eq:zero_intg}
	\chi_{ij}^{(b)}=-e^2\sum_n\frac{f_n}{\hbar^2}\dk\frac{\p^2\e_{n\bk}}{\p k_i\p k_j},
\end{equation}
the integral above is zero because of the translation symmetries of $\e_{n\bk}$ over the BZ. This is easily seen by doing the linear transformation $\bk=\sum x_j\bm{b}_j$, with $0\moi x_j\moi 1$. The integral in Eq.~\eqref{eq:zero_intg} can written as (apart from an unimportant constant Jacobian factor):
\begin{equation}
	\int_0^1 \hspace{-2mm} dx_1 \int_0^1 \hspace{-2mm} dx_2 \int_0^1 \hspace{-2mm} dx_3 \frac{\p^2\e_{n}(x_1\bm{b}_1+x_2\bm{b}_2+x_3\bm{b}_3)}{\p x_i\p x_j},
\end{equation}
where for clarity we have written the $\bk-$dependence in the energy bands as an argument instead of a subscript. If $i\neq j$ (say, 1 and 2), we have:
\begin{equation}
	\int_0^1 \hspace{-2mm} dx_3 [\e_n(\bm{b}_1+\bm{b}_2+x_3\bm{b}_3)-\e_n(x_3\bm{b}_3)],
\end{equation}
which is zero because of the periodicity of $\e_{n\bk}$ itself. If $i=j$ (say, 1), we have instead:
\begin{align}
	\int_0^1 \hspace{-2mm} dx_2 \int_0^1 \hspace{-2mm} dx_3 \Big[\frac{\p}{\p x_1}\e_n(\bm{b}_1+&x_2\bm{b}_2+x_3\bm{b}_3)-\notag\\
	&\frac{\p}{\p x_1} \e_n(x_2\bm{b}_2+x_3\bm{b}_3)\Big], 
\end{align}
which also vanishes by virtue of the periodicity of the $\bk-$derivatives of $\e_{n\bk}$. It must emphasized that the vanishing of the integral in Eq.~\eqref{eq:zero_intg} is allowed by the fact that the $f_n$ factors are $\bk$-independent (which permits to pull them out of the integral). Metals, where the Fermi level crosses an energy band, do not meet this requirement.

Thus, in Eq.~\eqref{eq:chi_S} only the term $\chil$ remains, given by Eq.~\eqref{eq:chi3}. The $f_{nn'}$ factors in these equations allow us to write the sum over valence ($v$) and conduction ($c$) states only. This results in:
\begin{align}\label{eq:chi4}
	\chil=\frac{e^2\hbar^2}{m^2}\dk \frac{1}{\e_{cv}^2}\Big[\frac{p_{j;cv}p_{i;vc}}{\e_{cv}-\hw-\ci\eta}+\notag\\
	\frac{p_{j;vc}p_{i;cv}}{\e_{cv}+\hw+\ci\eta}\Big],
\end{align}
where we have explicitly written the small imaginary part $\ci\eta$, {\it i.e.}, $\w$ is a real quantity in Eq.~\eqref{eq:chi4}. It must be noted that both Eq.~\eqref{eq:chi3} and its equivalent in Eq.~\eqref{eq:chi4} satisfy $\chil(-\w)=\chil(\w)^*$~\cite{LandauEOCM}. A further simplification can be achieved by using time reversal symmetry, which just amounts to the replacement $\bk\rightarrow-\bk$. After taking the imaginary part, Eq.~\eqref{eq:chi4} can be written as:
\begin{widetext}
\begin{align}\label{eq:chif}
	\text{Im}[\chil]=\frac{e^2\hbar^2\pi}{m^2}\dk\sum_{v,c}\frac{1}{\e_{cv}^2}\Big[\text{Re}[p_{i;vc}p_{j;cv}]\delta(\e_{cv}-\hw)-\text{Re}[p_{i;cv}p_{j;vc}]\delta(\e_{cv}+\hw) \Big],\;\w\in \mathbb{R}.
\end{align}
\end{widetext}
 In getting Eq.~\eqref{eq:chif} we have used:
\begin{equation}
	\frac{1}{\e_{nn'}-\hw\mp\ci\eta}=\mathcal{P}\frac{1}{\e_{nn'}-\hw}\pm \ci\pi\delta(\e_{nn'}-\hw),
\end{equation}
where $\mathcal{P}$ is the Cauchy principal value and all variables ($\e_{nn'}$, $\hw$ and $\eta$) are real. From Eq.~\eqref{eq:chif} it is straightforward to show that $\text{Im}[\chil(-\w)]=-\text{Im}[\chil(\w)]$, consistent with the more general condition $\chil(-\w)=\chil(\w)^*$ (with $\w$ real) mentioned above. It is this very condition that allows us to restrict the $\w$ range to $\w>0$ only. Also,  as $\e_{cv}=\e_c-\e_v$ is always positive, the restriction $\w>0$ amounts to dropping the last term in Eq.~\eqref{eq:chif}:
\begin{align}\label{eq:chiff}
	\text{Im}[\chil]&=\frac{e^2\hbar^2\pi}{m^2}\dk\sum_{v,c}\frac{1}{\e_{cv}^2}\times\notag\\
	&\text{Re}[p_{i;vc}p_{j;cv}]\delta(\e_{cv}-\hw) ,\;\w\in \mathbb{R}^+.
\end{align}
Since valence and conduction states are separated by an energy gap $\delta\e$ (in most semiconductors with practical applications this gap is the order of $1\,$eV), the difference $\e_{cv}$ is always positive and larger (or at most, equal) than $\delta\e$ and frequencies with $\hw<\delta\e$ gives $\text{Im}[\chil]=0$, without the need of evaluating the integral. This regular behavior at $\w=0$ is what one would expect for the static susceptibility of a semiconductor. (The finite value of the permittivity comes from the real part of $\chil$, which is nonzero even for $\hw$ within the energy gap.)

Before continuing, the following fact must be noted: In any numerical implementation, Eqs.~\eqref{eq:chif} and~\eqref{eq:chiff} cannot, of course, be calculated without resorting to {\it representations} of the Dirac delta. The most widely used are $\delta_{\eta}(\w)=(1/\pi)\eta/(x^2+\eta^2)$ (Poisson representation) and $\delta_\eta(\w)=e^{-\w^2/2\eta}/\sqrt{2\pi\eta}$ (Gaussian representation), or some variations of the latter using Hermite polynomials~\cite{Methfessel1989}. All of these approaches require a small positive parameter $\eta$ controlling the {\it broadening} of the representation [not necessarily equal to the small $\eta$ in Eq.~\eqref{eq:chi4}]. All of these approaches introduce a numerical error in the calculation of $\text{Im}[\chil]$, which is small when $\eta$ is in turn small. The optimal value of $\eta$ is intimately related to the sampling of the Brillouin zone, how fine or coarse it is, and in most cases has to be tuned {\it by hand} until the appropriate convergence is reached.

\subsection{The sum rule.}
To prove the TRK sum rule we start from Eq.~\eqref{eq:chif} and perform the integration  $\int_0^\infty d\w\,\w\,\text{Im}[\chil]$. The only $\w$ dependence is in the Dirac delta, so that $\int_0^\infty d\w\,\w\,\delta(\e_{cv}-\hw)=\e_{cv}/\hbar^2$, obtaining:
\begin{equation}\label{THK2}
	\int_0^\infty d\w\,\w\,\text{Im}[\chil]=\frac{e^2\pi}{m^2}\dk\sum_{v,c}\frac{\text{Re}[p_{i;vc}p_{j;cv}]}{\e_{cv}}.
\end{equation} 
On the other hand, we already know that $\chil^{(b)}$  in Eq.~\eqref{eq:chi2} is zero. The double sum in the second term can be decomposed into sums over valence and conduction states only, because the $f_{nn'}$ factors are zero for state pairs $(n,n')$ of the same kind (valence or conduction). By using $\bp_{vc}^*=\bp_{cv}$, and omitting the irrelevant factor $-e^2/m^2$, the equation $\chil^{(b)}=0$ can be reduced to:
\begin{align}
	\dk &\sum_n mf_n\delta_{ij}+\notag\\
	&2\dk\sum_{v,c}\frac{1}{\e_{vc}}\text{Re}[p_{j;vc}p_{i;cv}]=0,
\end{align}
but $\sum_nf_n=N_e$ is the number of valence electrons and the integral $\int d\bk$ is over the primitive cell in reciprocal space, thus equal to $(2\pi)^3/\Omega$. We  then get:
\begin{equation}
	\dk\sum_{v,c}\frac{1}{\e_{cv}}\text{Re}[p_{i;vc}p_{j;cv}]=\frac{m }{2}n_e\delta_{ij},
\end{equation}
where $n_e=N_e/\Omega$ (explicit $\bk$-dependence has been omitted). Thus, replacing in Eq.~\eqref{THK2} we finally get:
\begin{equation}\label{eq:TRKf}
	\int_0^\infty d\w\, \w\, \text{Im}[\chi_{ij}]=\frac{\pi}{2}\frac{n_e e^2}{m} \delta_{ij},
\end{equation}
which is the well known TRK sum rule for semiconductors. By using the relation between $\chil$ and the permittivity $\e_{ij}$, $\e_{ij}=1+4\pi\chil$ in Gaussian units, we have the better known relation~\cite{LandauEOCM}:
\begin{equation}\label{eq:TRKff}
	\int_0^\infty d\w\, \w\, \text{Im}[\e_{ij}]=\frac{2\pi^2e^2}{m} n_e\delta_{ij}.
\end{equation}
 As a particular case, when we work with cubic materials, it holds that $\chi_{ii}=\chi$ and $\chi_{ij}=0$ when $i\neq j$, and the expression above simply becomes $\int_0^\infty d\w\, \w\, \text{Im}[\epsilon]=2\pi^2e^2 n_e/m$.

From Eq.~\eqref{eq:TRKf} it can be noted that $\int_0^\infty d\w\, \w\, \text{Im}[\chi_{ij}^{(1)}]=0$ for $i\neq j$. However, this follows more straightforwardly from the symmetry of the material rather than the TRK sum rule. For instance, in many crystal classes~\cite{misc1} we have $\chil=0$ when $i\neq j$~\cite{Boyd,Shen}, making this case in Eq.~\eqref{eq:TRKf} sort of trivial. Also, the number $n_e$ in Eq.~\eqref{eq:TRKf} is the number of {\it valence} electrons per unit cell volume, as only these states contribute to the sum $\sum_n f_n$. This is the kind of weakly bound electron that enters in the high-frequency limit of $\chil$, as done in the literature~\cite{LandauSP2,CallawayQTOTSS,IbachSSP}.

\section{Conclusions}\label{sec:conclusions}
By using the periodicity of the $u_{n\bk}$ part of the Bloch eigenstates, we have obtained momentum matrix elements and the effective mass tensor that fully coincide with those obtained using the $\bk-$representation. These expressions  were then used to eliminate the singularity  of the linear susceptibility at $\w=0$, as commonly done in the literature. Remarkably, the same condition that makes the optical susceptibility non-singular at $\w=0$ leads to  the Thomas-Reiche-Kuhn sum rule. It must also be noticed that our results rely on the vanishing of $\chil^{(b)}$ in Eq.~\eqref{eq:chi2}, which happens only in the limit of a perfect semiconductor, where the integral of the effective mass tensor over the BZ also vanishes. All this treatment bypasses the high-frequency limit of $\chil$, which is the standard way of proving the TRK sum rule, and reinforces the intrinsically quantum mechanical origin of the TRK sum rule.

\section{Acknowledgements}
This work started as reflections during another, larger project on nonlinear optical properties in semiconductors funded by the Department of Energy.  My gratitude goes to them.

%


\begin{thebibliography}{22}%
\makeatletter
\providecommand \@ifxundefined [1]{%
 \@ifx{#1\undefined}
}%
\providecommand \@ifnum [1]{%
 \ifnum #1\expandafter \@firstoftwo
 \else \expandafter \@secondoftwo
 \fi
}%
\providecommand \@ifx [1]{%
 \ifx #1\expandafter \@firstoftwo
 \else \expandafter \@secondoftwo
 \fi
}%
\providecommand \natexlab [1]{#1}%
\providecommand \enquote  [1]{``#1''}%
\providecommand \bibnamefont  [1]{#1}%
\providecommand \bibfnamefont [1]{#1}%
\providecommand \citenamefont [1]{#1}%
\providecommand \href@noop [0]{\@secondoftwo}%
\providecommand \href [0]{\begingroup \@sanitize@url \@href}%
\providecommand \@href[1]{\@@startlink{#1}\@@href}%
\providecommand \@@href[1]{\endgroup#1\@@endlink}%
\providecommand \@sanitize@url [0]{\catcode `\\12\catcode `\$12\catcode `\&12\catcode `\#12\catcode `\^12\catcode `\_12\catcode `\%12\relax}%
\providecommand \@@startlink[1]{}%
\providecommand \@@endlink[0]{}%
\providecommand \url  [0]{\begingroup\@sanitize@url \@url }%
\providecommand \@url [1]{\endgroup\@href {#1}{\urlprefix }}%
\providecommand \urlprefix  [0]{URL }%
\providecommand \Eprint [0]{\href }%
\providecommand \doibase [0]{https://doi.org/}%
\providecommand \selectlanguage [0]{\@gobble}%
\providecommand \bibinfo  [0]{\@secondoftwo}%
\providecommand \bibfield  [0]{\@secondoftwo}%
\providecommand \translation [1]{[#1]}%
\providecommand \BibitemOpen [0]{}%
\providecommand \bibitemStop [0]{}%
\providecommand \bibitemNoStop [0]{.\EOS\space}%
\providecommand \EOS [0]{\spacefactor3000\relax}%
\providecommand \BibitemShut  [1]{\csname bibitem#1\endcsname}%
\let\auto@bib@innerbib\@empty
\bibitem [{\citenamefont {Sobota}\ \emph {et~al.}(2021)\citenamefont {Sobota}, \citenamefont {He},\ and\ \citenamefont {Shen}}]{Sobota2021}%
  \BibitemOpen
  \bibfield  {author} {\bibinfo {author} {\bibfnamefont {J.~A.}\ \bibnamefont {Sobota}}, \bibinfo {author} {\bibfnamefont {Y.}~\bibnamefont {He}},\ and\ \bibinfo {author} {\bibfnamefont {Z.-X.}\ \bibnamefont {Shen}},\ }\bibfield  {title} {\bibinfo {title} {Angle-resolved photoemission studies of quantum materials},\ }\href {https://doi.org/10.1103/RevModPhys.93.025006} {\bibfield  {journal} {\bibinfo  {journal} {Rev. Mod. Phys.}\ }\textbf {\bibinfo {volume} {93}},\ \bibinfo {pages} {025006} (\bibinfo {year} {2021})}\BibitemShut {NoStop}%
\bibitem [{\citenamefont {Barnett}\ and\ \citenamefont {Loudon}(2012)}]{Barnett2012}%
  \BibitemOpen
  \bibfield  {author} {\bibinfo {author} {\bibfnamefont {S.~M.}\ \bibnamefont {Barnett}}\ and\ \bibinfo {author} {\bibfnamefont {R.}~\bibnamefont {Loudon}},\ }\bibfield  {title} {\bibinfo {title} {Optical {T}homas-{R}eiche-{K}uhn {S}um {R}ules},\ }\href {https://doi.org/10.1103/PhysRevLett.108.013601} {\bibfield  {journal} {\bibinfo  {journal} {Phys. Rev. Lett.}\ }\textbf {\bibinfo {volume} {108}},\ \bibinfo {pages} {013601} (\bibinfo {year} {2012})}\BibitemShut {NoStop}%
\bibitem [{\citenamefont {Bransden}\ and\ \citenamefont {Joachain}(1983)}]{BransdenPAM}%
  \BibitemOpen
  \bibfield  {author} {\bibinfo {author} {\bibfnamefont {B.~H.}\ \bibnamefont {Bransden}}\ and\ \bibinfo {author} {\bibfnamefont {C.~J.}\ \bibnamefont {Joachain}},\ }\href@noop {} {\emph {\bibinfo {title} {Physics of atoms and molecules}}}\ (\bibinfo  {publisher} {Longman},\ \bibinfo {address} {Essex, England},\ \bibinfo {year} {1983})\ pp.\ \bibinfo {pages} {181--182}\BibitemShut {NoStop}%
\bibitem [{\citenamefont {Landau}\ and\ \citenamefont {Lifshitz}(1960)}]{LandauEOCM}%
  \BibitemOpen
  \bibfield  {author} {\bibinfo {author} {\bibfnamefont {L.~D.}\ \bibnamefont {Landau}}\ and\ \bibinfo {author} {\bibfnamefont {E.~M.}\ \bibnamefont {Lifshitz}},\ }\href@noop {} {\emph {\bibinfo {title} {Electrodynamics of Continuous Media}}}\ (\bibinfo  {publisher} {Pergamon Press},\ \bibinfo {address} {New York},\ \bibinfo {year} {1960})\BibitemShut {NoStop}%
\bibitem [{\citenamefont {Callaway}(1991)}]{CallawayQTOTSS}%
  \BibitemOpen
  \bibfield  {author} {\bibinfo {author} {\bibfnamefont {J.}~\bibnamefont {Callaway}},\ }\href@noop {} {\emph {\bibinfo {title} {Quantum Theory of the Solid State}}}\ (\bibinfo  {publisher} {Academic Press, Inc.},\ \bibinfo {address} {San Diego, CA.},\ \bibinfo {year} {1991})\BibitemShut {NoStop}%
\bibitem [{\citenamefont {Bassani}\ and\ \citenamefont {Scandolo}(1991)}]{Bassani1991}%
  \BibitemOpen
  \bibfield  {author} {\bibinfo {author} {\bibfnamefont {F.}~\bibnamefont {Bassani}}\ and\ \bibinfo {author} {\bibfnamefont {S.}~\bibnamefont {Scandolo}},\ }\bibfield  {title} {\bibinfo {title} {Dispersion relations and sum rules in nonlinear optics},\ }\href {https://doi.org/10.1103/PhysRevB.44.8446} {\bibfield  {journal} {\bibinfo  {journal} {Phys. Rev. B}\ }\textbf {\bibinfo {volume} {44}},\ \bibinfo {pages} {8446} (\bibinfo {year} {1991})}\BibitemShut {NoStop}%
\bibitem [{\citenamefont {Koutserimpas}\ \emph {et~al.}(2025)\citenamefont {Koutserimpas}, \citenamefont {Li}, \citenamefont {Miller},\ and\ \citenamefont {Monticone}}]{Koutserimpas2025}%
  \BibitemOpen
  \bibfield  {author} {\bibinfo {author} {\bibfnamefont {T.~T.}\ \bibnamefont {Koutserimpas}}, \bibinfo {author} {\bibfnamefont {H.}~\bibnamefont {Li}}, \bibinfo {author} {\bibfnamefont {O.~D.}\ \bibnamefont {Miller}},\ and\ \bibinfo {author} {\bibfnamefont {F.}~\bibnamefont {Monticone}},\ }\bibfield  {title} {\bibinfo {title} {Exploring the equivalence of causality-based and quantum mechanics-based sum rules for harmonic generation in nonlinear optical materials},\ }\href {https://doi.org/10.1364/OME.546776} {\bibfield  {journal} {\bibinfo  {journal} {Opt. Mater. Express}\ }\textbf {\bibinfo {volume} {15}},\ \bibinfo {pages} {413} (\bibinfo {year} {2025})}\BibitemShut {NoStop}%
\bibitem [{\citenamefont {Cabellos}\ \emph {et~al.}(2009)\citenamefont {Cabellos}, \citenamefont {Mendoza}, \citenamefont {Escobar}, \citenamefont {Nastos},\ and\ \citenamefont {Sipe}}]{Cabellos2009}%
  \BibitemOpen
  \bibfield  {author} {\bibinfo {author} {\bibfnamefont {J.~L.}\ \bibnamefont {Cabellos}}, \bibinfo {author} {\bibfnamefont {B.~S.}\ \bibnamefont {Mendoza}}, \bibinfo {author} {\bibfnamefont {M.~A.}\ \bibnamefont {Escobar}}, \bibinfo {author} {\bibfnamefont {F.}~\bibnamefont {Nastos}},\ and\ \bibinfo {author} {\bibfnamefont {J.~E.}\ \bibnamefont {Sipe}},\ }\bibfield  {title} {\bibinfo {title} {Effects of nonlocality on second-harmonic generation in bulk semiconductors},\ }\href {https://doi.org/10.1103/PhysRevB.80.155205} {\bibfield  {journal} {\bibinfo  {journal} {Phys. Rev. B}\ }\textbf {\bibinfo {volume} {80}},\ \bibinfo {pages} {155205} (\bibinfo {year} {2009})}\BibitemShut {NoStop}%
\bibitem [{\citenamefont {Ashcroft}\ and\ \citenamefont {Mermin}(1976)}]{Ashcroft76}%
  \BibitemOpen
  \bibfield  {author} {\bibinfo {author} {\bibfnamefont {N.~W.}\ \bibnamefont {Ashcroft}}\ and\ \bibinfo {author} {\bibfnamefont {N.~D.}\ \bibnamefont {Mermin}},\ }\href@noop {} {\emph {\bibinfo {title} {{S}olid {S}tate {P}hysics}}}\ (\bibinfo  {publisher} {Holt-Saunders},\ \bibinfo {year} {1976})\BibitemShut {NoStop}%
\bibitem [{\citenamefont {Blount}(1962)}]{BLOUNT1962305}%
  \BibitemOpen
  \bibfield  {author} {\bibinfo {author} {\bibfnamefont {E.}~\bibnamefont {Blount}},\ }\bibfield  {title} {\bibinfo {title} {Formalisms of band theory}\ }(\bibinfo  {publisher} {Academic Press},\ \bibinfo {year} {1962})\ pp.\ \bibinfo {pages} {305--373}\BibitemShut {NoStop}%
\bibitem [{\citenamefont {Landau}\ and\ \citenamefont {Lifshitz}(1980)}]{LandauSP2}%
  \BibitemOpen
  \bibfield  {author} {\bibinfo {author} {\bibfnamefont {L.~D.}\ \bibnamefont {Landau}}\ and\ \bibinfo {author} {\bibfnamefont {E.~M.}\ \bibnamefont {Lifshitz}},\ }\href@noop {} {\emph {\bibinfo {title} {Statistical Physics. Part 2}}}\ (\bibinfo  {publisher} {Pergamon Press},\ \bibinfo {address} {New York},\ \bibinfo {year} {1980})\BibitemShut {NoStop}%
\bibitem [{\citenamefont {Kittel}(1963)}]{KittelQTS}%
  \BibitemOpen
  \bibfield  {author} {\bibinfo {author} {\bibfnamefont {C.}~\bibnamefont {Kittel}},\ }\href@noop {} {\emph {\bibinfo {title} {Quantum Theory of Solids}}}\ (\bibinfo  {publisher} {Wiley},\ \bibinfo {address} {New York},\ \bibinfo {year} {1963})\BibitemShut {NoStop}%
\bibitem [{\citenamefont {Jackson}(1983)}]{JacksonCE}%
  \BibitemOpen
  \bibfield  {author} {\bibinfo {author} {\bibfnamefont {J.~D.}\ \bibnamefont {Jackson}},\ }\href@noop {} {\emph {\bibinfo {title} {Classical Electrodynamics}}},\ \bibinfo {edition} {3rd}\ ed.\ (\bibinfo  {publisher} {Wiley},\ \bibinfo {address} {New York, NY},\ \bibinfo {year} {1983})\BibitemShut {NoStop}%
\bibitem [{\citenamefont {Garc\'{i}a}\ and\ \citenamefont {Rodr\'{i}guez}(2023)}]{Garcia2023}%
  \BibitemOpen
  \bibfield  {author} {\bibinfo {author} {\bibfnamefont {J.~D.}\ \bibnamefont {Garc\'{i}a}}\ and\ \bibinfo {author} {\bibfnamefont {B.~A.}\ \bibnamefont {Rodr\'{i}guez}},\ }\bibfield  {title} {\bibinfo {title} {Quantum extension to the semiclassical theory of electrical susceptibility},\ }\href {https://doi.org/10.1364/JOSAB.502568} {\bibfield  {journal} {\bibinfo  {journal} {J. Opt. Soc. Am. B}\ }\textbf {\bibinfo {volume} {40}},\ \bibinfo {pages} {2999} (\bibinfo {year} {2023})}\BibitemShut {NoStop}%
\bibitem [{\citenamefont {Ziman}(1965)}]{ZimanPTS}%
  \BibitemOpen
  \bibfield  {author} {\bibinfo {author} {\bibfnamefont {J.~M.}\ \bibnamefont {Ziman}},\ }\href@noop {} {\emph {\bibinfo {title} {Principles of the theory of solids}}}\ (\bibinfo  {publisher} {Cambridge University Press},\ \bibinfo {address} {London},\ \bibinfo {year} {1965})\BibitemShut {NoStop}%
\bibitem [{\citenamefont {Sipe}\ and\ \citenamefont {Ghahramani}(1993)}]{Sipe1993}%
  \BibitemOpen
  \bibfield  {author} {\bibinfo {author} {\bibfnamefont {J.~E.}\ \bibnamefont {Sipe}}\ and\ \bibinfo {author} {\bibfnamefont {E.}~\bibnamefont {Ghahramani}},\ }\bibfield  {title} {\bibinfo {title} {Nonlinear optical response of semiconductors in the independent-particle approximation},\ }\href {https://doi.org/10.1103/PhysRevB.48.11705} {\bibfield  {journal} {\bibinfo  {journal} {Phys. Rev. B}\ }\textbf {\bibinfo {volume} {48}},\ \bibinfo {pages} {11705} (\bibinfo {year} {1993})}\BibitemShut {NoStop}%
\bibitem [{\citenamefont {Ghahramani}\ \emph {et~al.}(1991)\citenamefont {Ghahramani}, \citenamefont {Moss},\ and\ \citenamefont {Sipe}}]{Ghahramani1991}%
  \BibitemOpen
  \bibfield  {author} {\bibinfo {author} {\bibfnamefont {E.}~\bibnamefont {Ghahramani}}, \bibinfo {author} {\bibfnamefont {D.~J.}\ \bibnamefont {Moss}},\ and\ \bibinfo {author} {\bibfnamefont {J.~E.}\ \bibnamefont {Sipe}},\ }\bibfield  {title} {\bibinfo {title} {Full-band-structure calculation of second-harmonic generation in odd-period strained (\uppercase{S}i)$_n$/(\uppercase{G}e)$_n$superlattices},\ }\href {https://doi.org/10.1103/PhysRevB.43.8990} {\bibfield  {journal} {\bibinfo  {journal} {Phys. Rev. B}\ }\textbf {\bibinfo {volume} {43}},\ \bibinfo {pages} {8990} (\bibinfo {year} {1991})}\BibitemShut {NoStop}%
\bibitem [{\citenamefont {Methfessel}\ and\ \citenamefont {Paxton}(1989)}]{Methfessel1989}%
  \BibitemOpen
  \bibfield  {author} {\bibinfo {author} {\bibfnamefont {M.}~\bibnamefont {Methfessel}}\ and\ \bibinfo {author} {\bibfnamefont {A.~T.}\ \bibnamefont {Paxton}},\ }\bibfield  {title} {\bibinfo {title} {High-precision sampling for brillouin-zone integration in metals},\ }\href {https://doi.org/10.1103/PhysRevB.40.3616} {\bibfield  {journal} {\bibinfo  {journal} {Phys. Rev. B}\ }\textbf {\bibinfo {volume} {40}},\ \bibinfo {pages} {3616} (\bibinfo {year} {1989})}\BibitemShut {NoStop}%
\bibitem [{mis()}]{misc1}%
  \BibitemOpen
  \href@noop {} {}\bibinfo {note} {More precisely, this is true for the orthorhombic, tetragonal, trigonal, hexagonal and cubic crystal classes, as well as all isotropic materials.}\BibitemShut {Stop}%
\bibitem [{\citenamefont {Boyd}(2020)}]{Boyd}%
  \BibitemOpen
  \bibfield  {author} {\bibinfo {author} {\bibfnamefont {R.~W.}\ \bibnamefont {Boyd}},\ }\href@noop {} {\emph {\bibinfo {title} {Nonlinear optics}}},\ \bibinfo {edition} {4th}\ ed.\ (\bibinfo  {publisher} {Academic Press},\ \bibinfo {year} {2020})\BibitemShut {NoStop}%
\bibitem [{\citenamefont {Shen}(1984)}]{Shen}%
  \BibitemOpen
  \bibfield  {author} {\bibinfo {author} {\bibfnamefont {Y.~R.}\ \bibnamefont {Shen}},\ }\href@noop {} {\emph {\bibinfo {title} {The Principles of Nonlinear Optics}}}\ (\bibinfo  {publisher} {Wiley},\ \bibinfo {address} {New York},\ \bibinfo {year} {1984})\BibitemShut {NoStop}%
\bibitem [{\citenamefont {Ibach}\ and\ \citenamefont {L{\"u}th}(2009)}]{IbachSSP}%
  \BibitemOpen
  \bibfield  {author} {\bibinfo {author} {\bibfnamefont {H.}~\bibnamefont {Ibach}}\ and\ \bibinfo {author} {\bibfnamefont {H.}~\bibnamefont {L{\"u}th}},\ }\href@noop {} {\emph {\bibinfo {title} {Solid State Physics: An introduction to principles of material science}}}\ (\bibinfo  {publisher} {Springer},\ \bibinfo {address} {Berlin},\ \bibinfo {year} {2009})\BibitemShut {NoStop}%
\end{thebibliography}
\end{document}